\documentclass[11pt]{article}%{amsart}
\usepackage{amssymb}
\usepackage{latexsym}
\usepackage[mathscr]{eucal}
\usepackage{citesort}
\usepackage{url}
\newcommand{\ext}{{\mbox \scriptsize \rm ext}}
\newcommand{\Ric}{\mathrm{Ric}}
\newcounter{mnotecount}[section]

\newcommand{\mcM}{{\mycal M}}
\newcommand{\mcK}{{\mycal K}}
\newcommand{\mcL}{{\mycal L}}
\newcommand{\hyp}{{\mycal S}}

\def \Reel{\mathbb{R}}
\def \R {\Reel}

\newcommand{\be}{\begin{equation}}
\newcommand{\ee}{\end{equation}}
\newcommand{\bel}[1]{\begin{equation}\label{#1}}
\newcommand{\beal}[1]{\begin{eqnarray}\label{#1}}
\newcommand{\beadl}[1]{\begin{deqarr}\label{#1}}
\newcommand{\eeadl}[1]{\arrlabel{#1}\end{deqarr}}
\newcommand{\eeal}[1]{\label{#1}\end{eqnarray}}
\newcommand{\eead}[1]{\end{deqarr}}
\newcommand{\eea}{\end{eqnarray}}
\newcommand{\eeaa}{\end{eqnarray*}}

\newcommand{\Hess}{\mathrm{Hess}\,}

\newcommand{\eq}[1]{(\ref{#1})}

\DeclareFontFamily{OT1}{rsfs}{}
\DeclareFontShape{OT1}{rsfs}{m}{n}{ <-7> rsfs5 <7-10> rsfs7 <10->
rsfs10}{} \DeclareMathAlphabet{\mycal}{OT1}{rsfs}{m}{n}

\newcommand{\mcD}{{\mycal D}}

\newcommand{\qed}{\hfill $\Box$\bigskip}

\newcommand{\proof}{\noindent {\sc Proof:\ }}

\newtheorem{defi}{\sc Definition\rm}%[section]

\newtheorem{Theorem}[defi]{\sc Theorem\rm}

\newtheorem{Proposition}[defi]{\sc Proposition\rm}

\newtheorem{Corollary}[defi]{\sc Corollary\rm}

\newtheorem{Remark}[defi]{\sc Remark\rm}

\begin{document}
\title{A poor man's positive energy theorem}
\author{
Piotr T. Chru\'sciel\thanks{Partially supported by a Polish
Research Committee grant 2 P03B 073 24; email \protect\url{
piotr@gargan.math.univ-tours.fr}, URL \protect\url{
www.phys.univ-tours.fr/}$\sim$\protect\url{piotr}}\\ D\'epartement de
Math\'ematiques\\
Facult\'e des Sciences\\ Parc de Grandmont\\ F37200 Tours, France
\\
\\
Gregory J. Galloway\thanks{Partially supported by  NSF grant \#
DMS-0104042; email
     \protect\url{galloway@math.miami.edu}} \\Department of
Mathematics\\University of Miami\\Coral Gables, FL, 33124, USA
}
\date{}

\maketitle

\begin{abstract}
We show that positivity of energy for stationary, asymptotically
flat, non-singular domains of outer communications is a simple
corollary of  the Lorentzian splitting theorem.
\end{abstract}

A milestone in mathematical general relativity is the positive
energy theorem~\cite{SchoenYau81,Witten81}. It is somewhat
surprising that the standard proofs of this result do not have
anything to do with Lorentzian geometry. In this note
we prove energy positivity using purely Lorentzian techniques,
albeit for a rather restricted class of geometries; it seems that
in practice our proof only applies to stationary (with or without
black holes) space-times. This is a much weaker statement than the
theorems in~\cite{SchoenYau81,Witten81} and their various
extensions,
%(see~\cite{BartnikChrusciel1,Bray:preparation2,Herzlich97a,HI2,GHHP}),
but the proof here seems of interest because the techniques
involved are completely different and of a quite elementary
nature. Using arguments rather similar in spirit to those of the
classical singularity theorems \cite{HE}, our proof is a very
simple reduction of the problem to the Lorentzian splitting
theorem~\cite{galloway:complete}.  (In lieu of the Lorentzian
splitting theorem, one can impose the {\it generic condition}
\cite[p. 101]{HE}, thereby making the proof completely
elementary.) The approach taken here bares some relation to the
Penrose-Sorkin-Woolgar~\cite{PenroseSorkinWoolgar} argument for
positivity of mass, and indeed arose out of an interest in
understanding their work.

For $m\in\R$, let $g_m$ denote the $n+1$ dimensional, $n\ge 3$,
Schwarzschild metric with mass parameter $m$; in isotropic
coordinates \cite{Patel:1999ej}, \bel{stschw} g_m= \left(1 +
\frac{m}{2|x|^{n-2}}\right)^{\frac4{n-2}} \left(\sum_{1=1}^n
dx_i^2\right) -
\left(\frac{1-m/2|x|^{n-2}}{1+m/2|x|^{n-2}}\right)^2 dt^2\;. \ee
We shall say that a metric $g$ on $\R\times \left(\R^n\setminus
B(0,R)\right)$, $R^{n-2}>m/2$, is \emph{uniformly Schwarzchildian}
if, in the coordinates of \eq{stschw},
\bel{guS} g-g_m = o(|m|r^{-(n-2)})\;,\quad
\partial_\mu\left(g-g_m \right)= o(|m|r^{-(n-1)})\;.\ee
(Here $o$ is meant at fixed $g$ and $m$, with $r$ going to
infinity.) 
It is a
flagrant abuse of terminology to allow $m=0$ in this definition,
and we will happily abuse; what is meant in this case is that $g= g_0$, 
i.e., g is flat \footnote{The asymptotic conditions for the case $m=0$ of our
theorem are way too strong for a rigidity statement of real
interest, even within a stationary context. So it is fair to say
that our result only excludes $m<0$ for stationary space-times.},
for $r>R$.

Some comments about this notion are in order. First,  metrics as
above have constant Trautman-Bondi mass and therefore do not
contain gravitational radiation; one expects such metrics to be
stationary if physically reasonable field equations are imposed.
Next, every metric in space-time dimension four which is
stationary, asymptotically flat and vacuum or electro-vacuum in
the asymptotically flat region is uniformly Schwarzschildian there
when $m\ne 0$ (cf., e.g., \cite{Simon:1984kz}).

The hypotheses of our theorem below are compatible with stationary
black hole space-times with non-degenerate Killing horizons.

We say that the matter fields satisfy the \emph{timelike
convergence condition} if the Ricci tensor $R_{\mu\nu}$, as
expressed in terms of the matter energy-momentum tensor 
$T_{\mu\nu}$, satisfies the condition
\bel{tcc} R_{\mu\nu}X^\mu X^\nu \ge 0 \ \mbox{for all timelike
vectors $X^\mu$.}\ee

We define the \emph{domain of outer communications} of $\mcM$
 as the
intersection of the past $J^-(\mcM_\ext)$ of the asymptotic region 
$ \mcM_\ext = \R\times
\left(\R^n\setminus B(0,R)\right)$ with its future
$J^+(\mcM_\ext)$.

We need a version of weak asymptotic simplicity \cite{HE} for
asymptotically Schwarzschildian spacetimes.  We shall say that
such a spacetime  $(\mcM,g)$ is \emph{weakly asymptotically
regular}  if every null line starting in the domain of outer
communications (DOC) either crosses an event horizon (if any), or
reaches arbitrarily large values of $r$ in the asymptotically flat
regions.  By definition, a null line in $(\mcM,g)$ is an
inextendible null geodesic that is globally achronal; a timelike
line is an inextendible timelike geodesic, each segment of which
is maximal. Finally, we shall say that the DOC is {\it timelike
geodesically regular} if every timelike line in $\mcM$ which is
entirely contained in the DOC, and along which r is bounded, is
complete.

Our main result is the following: 

\begin{Theorem}
\label{Tpoorman} Let $(\mcM^{n+1}=\mcM,g)$ be an
$(n+1)$-dimensional space-time with matter fields satisfying the
timelike convergence condition, and suppose that $\mcM$ contains a
uniformly Schwarzschildian region \bel{mext} \mcM_\ext=\R\times
\left(\R^n\setminus B(0,R)\right)\;.\ee Assume that $(\mcM,g)$ is
weakly asymptotically regular and that the domain of outer
communications is timelike geodesically regular. If the domain of
outer communications of $\mcM$ has a Cauchy surface $\hyp$ which
is the union of one asymptotic end and of a compact interior
region (with a boundary lying at the intersection of the future
and past event horizons, if any), then $m>0$, unless $(\mcM,g)$
isometrically splits as $\R\times\hyp$ with metric
$g=-d\tau^2+\gamma$, $\mcL_{\partial_\tau}\gamma=0$, and
$(\hyp,\gamma)$ geodesically complete. Furthermore, the last case
does not occur if event horizons are present.
\end{Theorem}

   Before passing to the proof, we note the
following Corollary:

\begin{Corollary}\label{Cpoorman}
In addition to the hypotheses of Theorem~\ref{Tpoorman}, assume
that \bel{guS2} T_{\mu\nu}\in L^1\left(\R^n\setminus
B(0,R)\right)\;, \qquad
\partial_\nu
\partial_\mu g= O(r^{-\alpha})\;,\quad \alpha > 1+\frac n2\;.\ee
Then $m>0$ unless $\mcM$ is the Minkowski space-time.
\end{Corollary}

\noindent{\sc Proof of Theorem~\ref{Tpoorman}:} The idea is to
show that for $m\le 0$ the domain of outer communications contains
a timelike line, and the result then follows from the version of
the Lorentzian splitting theorem obtained
in~\cite{galloway:complete}. A straightforward computation shows
that  the Hessian $\Hess r=\nabla d r$ of $r$ is given by
\bel{Hessr2} \Hess r = -\frac{m}{r^{n-1}}
\left({(n-2)}dt^2 - dr^2 + r^2\,h\right) + r\,h + o(r^{-(n-1)})\;,
\ee where $h$ is the canonical metric on $S^n$, 
and the size of
the error terms refers to the components of the metric in
the coordinates of (\ref{stschw}).  Note that when $m<0$,
$\Hess r$, when restricted to the hypersurfaces of constant $r$,
is strictly positive definite for $r\ge R_1$, for some
sufficiently large $R_1$. Increasing $R_1$ if necessary, we can
obtain that $\partial_t$ is timelike for $r\ge R_1$. If $m=0$ we
set $R_1=R$. Let $p_{\pm k}$ denote the points $t=\pm k$, $\vec
x=(0,0,R_1)$; by global hyperbolicity there exists  a maximal
future directed timelike geodesic segment $\sigma_k$ from $p_{-k}$
to $p_{+k}$. We note, first, that the $\sigma_k$'s are obviously
contained in the domain of outer communications and therefore
cannot cross the black hole event horizons, if any. If $m=0$ then
$\sigma_k$ clearly cannot enter $\{r>R_1\}$, since timelike
geodesics in that region are straight lines which never leave that
region once they have entered it. It turns out that the same
occurs for $m<0$: suppose that $\sigma_k$ enters $\{r>R_1\}$, then
the function $r\circ \sigma_k$ has a maximum. However, if $s$ is
an affine parameter along $\sigma_k$ we have
$$ \frac {d^2 (r\circ \sigma_k)}{ds^2} = \Hess r (\dot \sigma_k,
\dot \sigma_k)>0$$ at the maximum if $m<0$, since
$dr(\dot\sigma_k)=0$ there, which is impossible. It follows that
all the $\sigma_k$'s (for $k$ sufficiently large) intersect the 
Cauchy surface $\hyp$ in the
compact set $\overline\hyp\setminus\{r>R_1\}$. A standard argument 
shows then that the $\sigma_k$'s accumulate to a timelike or null
line $\sigma$ through a point $p\in \overline\hyp$.  Let
$\{p_k\}=\sigma_k\cap \hyp$; suppose that $p\in \partial \hyp$,
then the portions of $\sigma_k$ to the past of $p_k$ accumulate at
a generator of the past event horizon $\dot
J^+\left(\mcM_\ext\right)$, and the portions of $\sigma_k$ to the
future of $p_k$ accumulate at a generator of the future event
horizon $\dot J^-\left(\mcM_\ext\right)$. This would result in
$\sigma$ being non-differentiable at $p$, contradicting the fact
that $\sigma$ is a geodesic. Thus the $p_k$'s stay away from
$\partial \hyp$, and $p\in \hyp$. By our ``weak asymptotic
regularity" hypothesis $\sigma$ cannot be null (as it does not
cross the event horizons, nor does it extend arbitrarily far  into
the asymptotic region). It follows that $\sigma$ is a timelike
line in $\mcM$ entirely contained in the globally hyperbolic
domain of outer communications $\mycal D$, with $r\circ \sigma$
bounded, and hence is complete by the assumed timelike geodesic
regularity of $\mcD$. Thus, one may apply~\cite{galloway:complete}
to conclude that $(\mycal D,g|_{\mycal D})$ is a metric product,
\bel{prod} g=-d\tau^2+\gamma\;,\ee for some $\tau$--independent
complete Riemannian metric $\gamma$. The completeness of this
metric product implies $\mycal D = \mcM$ (and in particular
excludes the existence of event horizons). \qed

\medskip

\noindent{\sc Proof of Corollary~\ref{Cpoorman}}: The lapse
function $N$ associated with a Killing vector field on a totally
geodesic hypersurface $\hyp$ with induced metric $\gamma$ and unit
normal $n$ satisfies the elliptic equation
$$\Delta_\gamma N - \mathrm{Ric}(n,n)N=0\;.$$
The vector field $\partial_\tau$ is a static Killing vector in
$\mcM_\ext$, and the usual analysis of groups of isometries of
asymptotically flat space-times shows that the metric $\gamma$ in
\eq{prod} is asymptotically flat. Again in \eq{prod} we have $N=1$
hence $\mathrm{Ric}(n,n)=0$, and the Komar mass of $\hyp$
vanishes.  By a theorem of Beig~\cite{BeigKomar} (originally
proved in dimension four, but the result generalises to any
dimensions under~\eq{guS2}) this implies the vanishing of the ADM
mass. Let $e_a$, $a=0,\ldots,n$, be an orthonormal frame with
$e_0=\partial_\tau$. The metric product structure implies that
$R_{0i}=0$. Thus, by the energy condition, for any fixed $i$ we
have
$$0\le \Ric(e_0+e_i,e_0+e_i)=R_{00}+R_{ii}=R_{ii}\;.$$
But again by the product structure, the components $R_{ii}$ of the
space-time Ricci tensor equal those of the Ricci tensor
$\Ric_\hyp$ of $\gamma$. It follows that $\Ric_\hyp \ge 0$. A
generalisation by Bartnik~\cite{Bartnik86} of
an argument of Witten~\cite{Witten81} shows that $(\hyp,\gamma)$
is isometric to Euclidean space; we reproduce the proof to make
clear its elementary character: Let $y^i$ be global harmonic
functions forming an asymptotically rectangular coordinate system
near infinity. Let $K^i =\nabla y^i$; then by Bochner's formula,
$$\Delta |K^i|^2
= 2|\nabla K^i|^2 + 2 \Ric_\hyp (K^i,K^i)\;.$$ Integrating the sum
over $i=1,\ldots,n$ of this gives the ADM mass as boundary term at
infinity, and the $\nabla y^i$ are all parallel. Since $\hyp$ is
simply connected at infinity, it must be Euclidean space. \qed

We close this note by showing  that the conditions on geodesics in
Theorem~\ref{Tpoorman} are always satisfied in  stationary domains
of outer communications.

\begin{Proposition} \label{Pstatwas} Let
the domain of outer communications $\mcD$ of $(\mcM,g)$ be
globally hyperbolic, with a Cauchy surface $\hyp$ such that
$\overline \hyp$   is the union of a finite number of
asymptotically flat regions and of a compact set  (with a boundary
lying at the intersection of the future and past event horizons,
if any). Suppose that there exists on $\mcM$ a Killing vector
field $X$ with complete orbits which is timelike, or
stationary-rotating\footnote{See~\cite{ChWald1} for the
definition.} in the asymptotically flat regions. Then the weak
asymptotic regularity and the timelike regularity
 conditions hold.
\end{Proposition}

\begin{Remark}
\label{Rstatwas} {\em We note that there might exist maximally
extended null geodesics in $(\mcD,g)$ which are trapped in space
within a compact set (as happens for the Schwarzschild metric),
but those geodesics will not be achronal.}
\end{Remark}

\proof
  By~\cite[Propositions~4.1
and 4.2]{ChWald1} we have $\mcD=\R\times \hyp$, with the flow of
$X$ consisting of translations along the $\R$ axis:
\bel{gstat}g=\alpha d\tau^2 + 2
\beta d\tau + \gamma\;, \quad X=\partial_\tau\;,
%\quad \mcL_X \beta=\mcL_X \gamma=0\;,,
\ee where $\gamma$ is a Riemannian metric on $\hyp$ and $\beta$ is
a one-form on $\hyp$. (We emphasise that we do not assume $X$ to
be timelike, so that $\alpha=g(X,X)$ can change sign.) Let
$\phi_t$ denote the flow of $X$ and let
$\sigma(s)=(\tau(s),p(s))\in \R\times \hyp$ be an affinely
parameterised causal line in $\mcD$, then for each $t\in\R$ the
curve $\phi_t(\sigma(s))=(\tau(s)+t,p(s))$ is also an affinely
parameterised causal line in $\mcD$. Suppose that there exists a
sequence $s_i$ such that $p(s_i)\to \partial \hyp$, setting
$t_i=-\tau(s_i)$ we have $\tau(\phi_{t_i}(\sigma(s_i))=0$, then
the points $\{p_{k_i}\}=\phi_{t_i}(\sigma)\cap \hyp$ accumulate at
$\partial \hyp$, which is not possible as in the proof of
Theorem~\ref{Tpoorman}. Therefore there exists an open
neighborhood $\mcK$ of $\partial \hyp$ such that $\sigma
\cap\left(  \R\times \mcK\right)=\emptyset$. This implies in turn
that $\sigma$ meets all the level sets of $\tau$. Standard
considerations using the fact that $\mcD$ is a stationary, or
stationary-rotating domain of outer communications (cf.,
e.g.,~\cite{ChWald1}) show that for every $p,q\in \hyp$ there
exists $T>0$ and a timelike curve from $(0,p)$ to $(T,q)$. The
constant $T$ can be chosen independently of $p$ and $q$ within the
compact set $\hyp\setminus\left(\mcK\cup\{r\ge R_1\}\right)$, with
$R_1=\sup_\sigma r$. It follows that an inextendible null geodesic
which is bounded in space within a compact set cannot be achronal,
so that $\sigma$ has to reach arbitrarily large values of $r$, and
weak asymptotic regularity follows. Similarly, if $\sigma$ is a
timelike line bounded in space within a compact set, then there
exists $s_1>0$ such that for any point $(\tau(s),p(s))$ with
$s=s_1+u$, $u>0$ one can find a timelike curve from $(0,p(0))$ to
$(\tau(s),p(s))$ by going  to the asymptotic region, staying there
for a time $u$, and coming back. The resulting curve will have
Lorentzian length larger than $u/2$ if one went sufficiently far
into the asymptotic region, and since $\sigma$ is
length-maximising it must be complete.
\qed

The key point in the argument above is non-existence of observer
horizons contained in the DOC. Somewhat more generally, we have
the following result, which does not assume existence of a Killing
vector:

\begin{Proposition}
\label{Pnoobh} Suppose that causal lines $\sigma$, with $r\circ
\sigma$ bounded, and which are contained entirely in $\mcD$,  do
not have \emph{observer horizons} extending to the asymptotic
region $\mcM_\ext$ (see \eq{mext}): \bel{noobh}\dot
J^{\pm}(\sigma;\mcD)\cap \mcM_\ext=\emptyset\;.\ee Then the weak
asymptotic regularity and the timelike regularity conditions hold.
\end{Proposition}

\proof It follows from \eq{noobh} that for any $u>0$ and for any
$s_1$ there exists $s_2$ and  a timelike curve $\Gamma_{u,s_1}$
from $\sigma(s_1)$ to $\sigma(s_2)$ which is obtained by following
a timelike curve from $\sigma(s_1)$  to the asymptotic region,
then staying there at fixed space coordinate for a coordinate time
$u$, and returning back to $\sigma$ along a timelike curve. One
concludes as in the proof of Proposition~\ref{Pstatwas}.
\qed

\smallskip

~

  \noindent {\sc Acknowledgements} PTC acknowledges the friendly
hospitality and financial support from the Department of
Mathematics of the  University of Miami during part of work on
this paper. We are grateful to Robert Bartnik for useful comments.

\bibliographystyle{amsplain}

\def\cprime{$'$}
\providecommand{\bysame}{\leavevmode\hbox to3em{\hrulefill}\thinspace}
\providecommand{\MR}{\relax\ifhmode\unskip\space\fi MR }
\providecommand{\MRhref}[2]{%
  \href{http://www.ams.org/mathscinet-getitem?mr=#1}{#2}
}
\providecommand{\href}[2]{#2}

\end{document}